\documentstyle[twocolumn,prl,aps, graphicx, epsf]{revtex}
\input{epsf}
\def\be{\begin{equation}}
\def\ee{\end{equation}}
\def\bea{\begin{eqnarray}}
\def\eea{\end{eqnarray}}
\begin{document}
\title
{Polarization transfer in the  $^{16}$O$(\vec e,e^{\prime} \vec p)
^{15}$N reaction}
\author{
S.~Malov,$^{25}$
K.~Wijesooriya,$^{33}$
F.~T.~Baker,$^{7}$
L.~Bimbot,$^{22}$
E.~J.~Brash,$^{24}$ 
C.~C.~Chang,$^{16}$ 
J.~M.~Finn,$^{33}$ 
K.~G.~Fissum,$^{17}$ 
J.~Gao,$^{17}$ 
R.~Gilman,$^{25,30}$
% ...
}
\address{
{\rm (The Jefferson Lab Hall A Collaboration)}\\
\vspace{0.1cm}
% ...
$^{7}$University of Georgia, Athens, Georgia, 30602\\
% ...
$^{16}$University of Maryland, College Park, Maryland, 20742\\
$^{17}$Massachusetts Institute of Technology, Cambridge, Massachusetts, 02139\\
% ...
$^{22}$Institut de Physique Nucl\'eaire, F-91406 Orsay, France\\
% ...
$^{24}$University of Regina, Regina, Saskatchewan, Canada, S4S 0A2\\
$^{25}$Rutgers, The State University of New Jersey, Piscataway, New Jersey, 
08854\\
% ...
$^{30}$Thomas Jefferson National Accelerator Facility, Newport News, Virginia, 
23606\\
% ...
$^{33}$College of William and Mary, Williamsburg, Virginia, 23187\\
% ...
}
\vspace{0.1cm}
\date{\today}
\maketitle
\begin{abstract}
The first $(\vec e,e^{\prime} \vec p)$ polarization transfer measurements
on a nucleus heavier than deuterium have been carried out at
Jefferson Laboratory. Transverse and longitudinal components of the 
polarization of protons ejected in the reaction 
$^{16}$O$(\vec e,e^{\prime} \vec p\,)$ were measured in 
quasielastic perpendicular kinematics at a $Q^2$ of 0.8 (GeV/c)$^2$. The data
are in good agreement with state of the art calculations.
\end{abstract}

\vspace{0.2cm}
\vspace{0.35cm}

Polarization transfer in the $(\vec e,e^{\prime}\vec p\,)$ 
reaction on a proton target is a direct measure of the ratio of the 
electric and magnetic form factors of the proton, $G_E^p/G_M^p$. 
When such measurements are carried out
on a nuclear target, the polarization transfer observables are sensitive 
to the form factor ratio of the proton embedded in the nuclear medium.
Because such experiments involve the measurement of ratios of polarizations 
at a single kinematic setting, the systematic errors are small, and different 
from those in standard Rosenbluth separations.

 We report here
 measurements of polarization transfer in the
 $^{16}$O$(\vec e,e^{\prime}\vec p\,)^{15}$N 
reaction, the first such measurement on a nucleus
 heavier than deuterium~\cite{Milbrath}.
The experiment, E89-033 at the Thomas Jefferson 
National Accelerator Facility (JLab),
was part of the commissioning effort for Hall A~\cite{Glas}. It was the 
first experiment to use polarized beam at JLab, and the first to use the 
focal plane polarimeter (FPP) mounted on the high-resolution hadron 
spectrometer.
Comparison of the results 
with state of the art calculations lays the groundwork for 
 high precision tests of changes of the form factors in the nuclear medium. 
The distorted-wave impulse approximation
(DWIA) provides a good description of the reaction, and 
the predictions are shown to be insensitive to 
various theoretical corrections.

The issue of possible modification of the properties of hadrons in the 
nucleus has been attracting experimental and theoretical attention for 
some years. It remains unsettled. 
Interpretations of inclusive $(e,e^{\prime})$ cross-section 
measurements in the y-scaling regime suggest that the radius of the nucleon 
is changed by less than a few percent at least for
 values of the four-momentum transfer squared (Q$^2$) up 
to about 1~(GeV/c)$^2$\cite{Sick}. 
These measurements are primarily sensitive to the 
magnetic form factor. Measurements of the Coulomb sum rule over a similar 
region in Q$^2$ indicate that the electric form factor in $^3$He 
is close to its free value~\cite{Schia},
 and some studies suggest that this is true in 
$^{12}$C and $^{56}$Fe as well~\cite{Jourdan}, but recent work disputes
this conclusion~\cite{Meziani}. 
Attempts to measure the ratio of electric to magnetic form factors 
of the nucleon in nuclei by Rosenbluth separations 
of cross sections in $(e,e^\prime p)$ reactions 
indicated  changes of  about 25\% at low Q$^2$ \cite{Rosen}, 
but some other experiments and theoretical analyses disagree~\cite{Van}. 
Recent theoretical work by the Adelaide group based on the quark-meson
coupling model predicted changes in the ratio of the two form 
factors for $^{16}$O of
roughly 10\% at Q$^2$ around 1~(GeV/c)$^2$ and about  
20\% or larger at about 2.5~(GeV/c)$^2$ \cite{Adelaide}.  Changes of 
this magnitude have also been suggested previously~\cite{Miller}.

For the free nucleon, the polarization transfer can be written in terms of the
form factors as~\cite{Arnold} 
$$I_0 P_l^\prime = {{E+E^\prime}\over{m_p}}\sqrt{\tau(1+\tau)}
G_M^2\tan^2(\theta/2)$$
$$I_0 P_t^\prime = -2\sqrt{\tau(1+\tau)}G_M G_E \tan(\theta/2)$$
$$I_0 = G_E^2 +\tau G_M^2 [1+2(1+\tau)\tan^2(\theta/2)]$$
$$\tau = Q^2/4m_p^2,$$

\noindent where $E$ and $E^\prime$ are the energies of the incident and scattered 
electron, $\theta$ is the electron scattering angle, and $m_p$ is
the proton mass.  $P_l^\prime$ and $P_t^\prime$ are the longitudinal 
and transverse polarization transfer observables, respectively.
The two components of the actual polarization in the scattering
plane~\cite{Raskin} are $hP_l^\prime$, parallel to the proton momentum, and 
$hP_t^\prime$, perpendicular to the proton momentum;
 $h$ is the electron beam polarization. The 
measured polarizations change sign 
when the electron helicity changes sign, so these polarization transfer 
quantities are insensitive to instrumental asymmetries in the detectors.

The ratio of the transferred polarizations is then
$${{P_t^\prime}\over{P_l^\prime}} =
{{-2m_p}\over{(E+E^\prime )\tan(\theta/2)}}{G_E\over{G_M}}.$$  
For a free proton target, 
the ratio of polarizations can be used to determine the ratio
of the form factors with small systematic errors; systematic problems
associated with Rosenbluth separations are eliminated.
The ratio is  independent of the beam polarization (assuming it is not zero)
and of the analyzing power of the proton polarimeter. 
One experimental datum requires a coincidence measurement at a single 
kinematic setting. 
The systematic error on the 
ratio of polarizations in the present experiment is  about $\pm$0.022,
due almost entirely to uncertainty in the precession of the proton's 
spin in the hadron spectrometer.
Even smaller errors have been achieved
in the subsequent E93-027  measurements
of the free form-factor ratio on a liquid hydrogen target at a Q$^2$ 
of 0.79~(GeV/c)$^2$ \cite{Perd}. 

For nuclear targets, the polarization transfer observables  
depend sensitively on the nucleon form factors, but they 
depend also on the nuclear wave functions.  In addition, they 
may be affected by final-state interactions of the outgoing proton,  
meson exchange and isobar currents, off-shell effects and the distortion of spinors
by strong Lorentz scalar and vector potentials~\cite{Boffi,Kelly1,Marco,Ryck,Kelly2,Udias}. 

Electrons from the CEBAF acelerator of 
energy 2.45 GeV and longitudinal polarization about 30\% were focussed on a
waterfall target with  three foils whose total
thickness was about $0.39$~g/cm$^2$. Scattered electrons were detected
in the focal plane array of the high resolution electron spectrometer 
in Hall A at a fixed laboratory angle of 
23.4$^\circ$ and a fixed central energy
of 2.00~GeV, the quasielastic peak.  Protons with a fixed 
central momentum of 973~MeV/c were 
detected in coincidence with electrons in the focal plane array of the 
hadron spectrometer. Measurements in quasi-perpendicular kinematics were made
at proton angles of  53.3$^\circ$ (for hydrogen data only), 55.7$^\circ$, and 
60.5$^\circ$, corresponding
to central missing momenta $p_m$ of 0, 85, and 140 MeV/c. 
Elastic scattering from hydrogen dominates the spectrum at  
53.3$^\circ$ 
and is visible also at 55.7$^\circ$.
The missing-mass  resolution of about 1 MeV was sufficient to 
easily distinguish the p$_{1/2}$ ground
state of $^{15}$N from the strongly excited  
p$_{3/2}$ state at 6.32 MeV, but small contributions from nearby weakly 
excited states could not be entirely excluded. 
 In the continuum, a peak corresponding primarily to knockout of nucleons 
from the s$_{1/2}$ shell rises weakly 
above a (physics) background presumably related to multi-particle 
knockout. 

The JLab focal plane polarimeter (FPP) was designed and built by a
collaboration of Rutgers, William \& Mary, Georgia,
Norfolk State, and Regina~\cite{fpp,Krishni,Malov}. The polarimeter, 
consisting 
of four tracking straw chambers and a graphite analyzer set to a thickness 
of  22.5~cm for this experiment, is mounted in the hadron spectrometer behind
vertical drift chambers and scintillators in the focal plane.
The analyzing power $A_c$  of the FPP  was taken from
the parametrization by McNaughton~{\it et al.}~\cite{Anal}. Measured values
of $A_c$, obtained by analyzing data from scattering on hydrogen, have 
been shown to agree well with this parameterization\cite{Perd}. The 
beam polarization was measured at varying intervals with a  
Mott polarimeter in the injector beam line.  For the 85~MeV/c data point
on $^{16}$O, values of the beam polarization 
determined from the Mott polarimeter and from the FPP results for hydrogen 
are in good agreement, well within the 5\% systematic error assigned 
to the beam polarization in the subsequent analysis of the oxygen data.
The  result for the ratio $\mu{G_E\over{G_M}}$ of  
hydrogen measured in this experiment at a 
Q$^2$ of 0.8 (GeV/c)$^2$ is  0.92$\pm$0.05, in agreement with 
previous results and with the values subsequently measured  
with higher precision~\cite{Perd}. 

\begin{figure}
  \centering\includegraphics[width=3.5in]{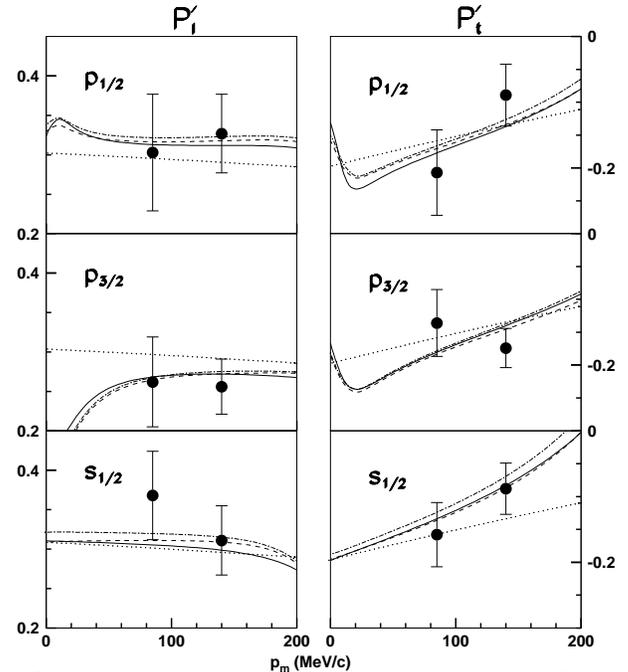}
  \caption[]{Measured values of the polarization transfer observables
	$P_l^{\prime}$ and $P_t^{\prime}$ for the
	$^{16}$O$(\vec e,e^{\prime}\vec p)^{15}$N reaction at
	$Q^2$ = 0.8 (GeV/c)$^2$. The theoretical curves represent 
	plane-wave calculations (dotted) and distorted-wave 
	calculations without spinor distortions (dashed)
	and with spinor distortions (dash-dot) by Kelly~\cite{Kelly2} 
	and by Udias {\it et al.}~\cite{Udias} (solid).}
	\label{M1}
\end{figure}

Results for the transverse and longitudinal components of the 
polarization at the two central values of the missing momentum 
for the two bound states, p$_{1/2}$ and p$_{3/2}$, and for the region 
of the unbound s$_{1/2}$ state are
shown in Fig.~\ref{M1}. The missing energy cuts on the latter 
were about 26 MeV wide. The polarizations are given in the 
scattering (lab) frame, defined by the incident and outgoing 
electron~\cite{Raskin,Kelly1}.
The errors shown are statistical. Systematic 
errors on the individual polarizations are about $\pm$6\%, primarily
due to the uncertainty in the polarization of the beam.
Small acceptance averaging corrections are 
included~\cite{Krishni}, as are the effects of corrections to the dipole 
approximation for spin transport of the proton in the hadron
spectrometer~\cite{Krishni,Malov}. Both corrections are generally  
less than about 2\%. Radiative corrections, expected to be much smaller 
than the statistical errors here, have not been made~\cite {Afanasev}. 

The values of $P_l^{\prime}$ and $P_t^{\prime}$ for
 the free proton measured in this 
experiment (via the hydrogen in the waterfall target) 
are $0.30\pm 0.01$ and $-0.20\pm 0.01$, with  
statistical errors. 
All the results for $P_l^\prime$ 
for $^{16}$O are within about one standard deviation
 of the free values, even those for the $s_{1/2}$ region; their 
average value is 0.30. Several of the $P_t^\prime$ data points deviate 
somewhat from the free value; their average is  
 -0.17. Such differences are not  
unexpected, even in the plane-wave impulse approximation (PWIA).

The curves in Fig.~\ref{M1} represent theoretical calculations
based upon one-body currents and free (MMD)~\cite{MMD} 
proton form factors.  PWIA
results, shown as dotted lines, are identical for the three states
and, at $p_m=0$, are equal to those for the free proton~\cite{Kelly1}. 
Final-state
interactions included in DWIA
calculations produce small state-dependent deviations from PWIA. The 
DWIA calculations by Kelly~\cite{Kelly1}  are based upon a relativized
Schr\"{o}dinger equation and an effective momentum approximation (EMA)
to the current operator.
The dashed curves assume that lower and upper components of bound and
ejectile spinors are related in the same way as for free protons \cite{Kelly1}.
The  dash-dotted curves include relativistic dynamics (spinor distortions) 
through the
effect of Dirac scalar and vector potentials upon the effective 
current operator~\cite{Kelly2}. 
 The solid curves show the results of 
 calculations by Moya de Guerra and Udias \cite{Udias} who 
 solve the Dirac equation directly without using the EMA.
All DWIA calculations shown used the same input as 
 the calculations of unpolarized observables in Ref. \cite{Gao}.
These include the 
EDIAO optical model of Cooper {\it et al.}~\cite{Cooper},
 NLSH bound-state wave functions~\cite{NLSH},
the Coulomb gauge, and  the cc2 off-shell current operator.
For modest $p_m$, the recoil polarization is relatively insensitive
to variations of these choices. 

\begin{figure}
  \begin{minipage}[t]{1.\linewidth}
  \centering \includegraphics[width=3.5in]{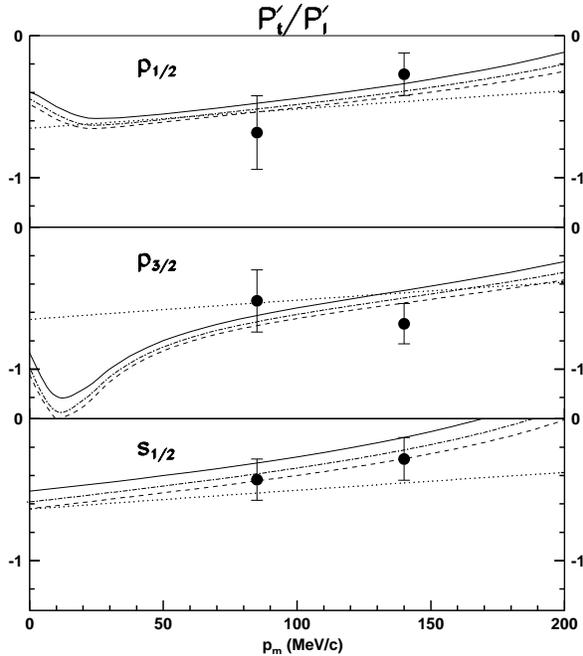}
  \caption[]{Measured values of the ratio of polarization transfer observables
	$P_t^\prime$/ $P_l^\prime$
	for the $^{16}$O$(\vec e,e^\prime\vec p)^{15}$N reaction at
	$Q^2$ = 0.8 (GeV/c$)^2$. The theoretical curves represent  
	plane-wave calculations (dotted) and distorted-wave  calculations  
	without spinor distortions (dashed) and with  spinor 
	distortions (dash-dotted)~\cite{Kelly2}. 
	The solid curves include the modifications of the nucleon 
	predicted by Lu {\it et al.}~\cite{Adelaide}   
	and spinor distortions.}
	\label{M2}
  \end{minipage}
\end{figure}

All DWIA calculations are in reasonable agreement with the measured data.
The two  calculations which include 
relativistic dynamics are very similar over the 
relevant range of $p_m$. This is expected, since the results of 
 the calculation of unpolarized observables in Ref.~\cite{Gao} suggested that 
  Kelly's formulation 
is a good approximation to the more accurate approach of Udias
{\it et al.} for $p_m \lesssim 300$ MeV/c.    
The effects of relativity on the recoil polarization 
are small for this range of $p_m$ and are dominated by distortion
of the ejectile spinor. 

Contributions 
from meson exchange (MEC) and isobar (IC) currents can also affect 
the recoil polarizations.  Early calculations of the effects of MEC and IC 
were carried out by the Pavia group\cite{Boffi}, and   
preliminary calculations by M. Radici~\cite{Marco} of this group 
for the present kinematics have been made. Predictions of these 
effects using a different approach have been published recently by  
J.~Ryckebusch {\it et al.}~\cite{Ryck} for the present kinematics.
The scale of these effects is typically comparable to the differences 
among the three DWIA curves shown. The DWIA with small corrections thus 
provides a firm baseline for considering changes in the form factor.

The ratios of the $P_t^\prime$ to $P_l^\prime$ data 
for the three states are plotted in  Fig.~\ref{M2}. 
Three of the theoretical curves shown there 
correspond to those in Fig.~\ref{M1}, namely the PWIA (dots), and 
the DWIA without spinor distortion (dashes) and with spinor distortion
(dash-dot) by Kelly~\cite{Kelly2}.
The data are in good agreement with the three predictions,
 as expected from Fig.~\ref{M1}. The ratio of the experimental and 
theoretical values of $P_t^\prime$/ $P_l^\prime$ for the summed p state
data is 0.95$\pm$ 0.18 using either DWIA calculation.

Deviations from unity significantly outside theoretical and experimental 
uncertainties would be evidence for changes in the nucleon form 
factor ratio in the nuclear medium. As noted in the introduction, the 
Adelaide group~\cite{Adelaide} 
obtained density-dependent form factors using a quark-meson coupling 
model and found changes in the  form factor ratio for $^{16}$O of about 
10\% for Q$^2$ = 0.8 (GeV/c)$^2$. The sensitivity of the (e, e$^{\prime}$ p) 
reaction to such changes has been estimated by Kelly using a
 local density approximation to the current operator.   
The fourth curve (solid) in Fig.~\ref{M2} shows that 
the 10\% changes in the form factors 
 translate into  changes of roughly  5\%   
in the  $P_t^\prime$ to $P_l^\prime$ ratio~\cite{Kelly2}.
 The reduced sensitivity of knockout at small p$_m$ can be understood 
by comparing the averaging procedure used by Lu {\it et al.} 
with one more closely related to the matrix elements involved in the 
 (e,e$^{\prime}$p) reaction.

Lu {\it et al.}~\cite{Adelaide} estimated the effect of density dependence upon
the electromagnetic form factors for a bound nucleon in
 orbital $\phi_{\alpha}$
 by using average form factors of the form
\begin{equation}
\bar{G}_\alpha(Q^2) \propto \int d^3r \; w_\alpha(r) G(Q^2,\rho_B(r))
\end{equation}
where 
 $\rho_B$ is the ground-state baryon density for the residual nucleus.
Here proportionality denotes division by a similar integral omitting $G$.
The static weighting factor 
$w_\alpha(r) = | \phi_\alpha(r) |^2$
 determines the effective density for different orbits.
In the (e, e$^{\prime}$ p) reaction, however, 
Kelly~\cite{Kelly2} finds that the 
 weighting factor is approximately 
\begin{equation}
w_\alpha =   \exp{( i \bbox{q} \cdot {\bf r} )} 
  \chi^{(-)}({\bf p}^\prime,{\bf r})^\ast    
  \phi_\alpha({\bf r})
\end{equation}
where $\chi$ is the distorted wave for ejectile momentum ${\bf p}^\prime$,
${\bf q}$ is the momentum transfer, and ${\bf p}_m = {\bf p}^\prime - {\bf q}$.
In the interests of simplicity, recoil corrections and details of the 
current operator have been suppressed.
In the plane-wave approximation, the weighting factor becomes 
\begin{equation}
w_\alpha^{\rm (PWIA)} = 
\exp{( -i {\bf p}_m \cdot {\bf r} )}\phi_\alpha({\bf r})   
\end{equation}
and for $p_m \rightarrow 0$ reduces to simply $\phi_\alpha$. 
In kinematic regimes explored thus far, 
 this linear dependence upon $\phi_\alpha$
 reduces the effect of 
density dependence in the reaction,
although the effect does increase with $p_m$.
Furthermore, absorption and nonlocality corrections also reduce interior
contributions to the average form factor.  

Much more precise measurements of the $P_t^\prime/P_l^\prime$ ratio
are now possible. 
The polarized beam at JLab has shown  a marked improvement in 
intensity, lifetime, and polarization since the commissioning experiment
so that the statistical errors 
can be greatly reduced within reasonable running times, 
and systematic errors are already small.
Conditions at the MAMI accelerator 
at Mainz are also appropriate for a high precision measurement at 
low Q$^2$, and an experiment was recently carried out 
there on $^4$He~\cite{Ransome}. However, although recoil 
polarization provides a direct signal for medium modifications of
nucleon form factors, the effect in (e,e$^{\prime}$p) reactions is
 smaller than previously expected. A rigorous interpretation will
 require a unified relativistic treatment of the reaction
 and form factor models, including two-body currents.

The present experiment  confirms 
the accuracy of the DWIA description of the reaction 
mechanism in this kinematical regime. The measured 
ratio of the transverse to longitudinal polarization 
transfers for the proton embedded in $^{16}$O at a Q$^2$
of 0.8~(GeV/c)$^2$ is in good agreement 
with calculations based on the free proton form factor with an 
experimental  uncertainty of about 18\%. 
The current generation of polarization transfer experiments 
should substantially improve this limit, but reliable identification of 
changes in the form factor in the medium remains an ambitious undertaking.

We are grateful to the entire technical staff at JLab for their 
dedicated and expert support of the commissioning experiment. We thank 
E. Moya De Guerra, M. Radici, J. Ryckebusch and J. M. Udias for numerous 
discussions and detailed calculations.  
 This work was supported in part by the National Science Foundation and 
the Department of Energy (U.S.), the National Institute for Nuclear Physics (Italy),
the Atomic Energy Commission and the National Center for Scientific 
Research (France), and the Natural Sciences and Engineering Research 
Council (Canada).

\end{document}